\documentclass[twocolumn,aps,prd,epsecnum,nofootinbib,10pt,preprintnumbers]{revtex4}
\usepackage{graphicx}
\usepackage{amssymb}

\newcommand{\beq}{\begin{equation}}
\newcommand{\eeq}{\end{equation}}
\newcommand{\beqa}{\begin{eqnarray}}
\newcommand{\eeqa}{\end{eqnarray}}

\begin{document}
\title{Supersymmetric heterotic solutions via non-$SU(3)$ standard embedding}

\author{
${}^{1}$Kazuki Hinoue\footnote{E-mail: hinoue@sci.osaka-cu.ac.jp},
${}^{2}$Shun'ya Mizoguchi\footnote{E-mail: mizoguch@post.kek.jp},
${}^{1}$Yukinori Yasui\footnote{E-mail: yasui@sci.osaka-cu.ac.jp}
}
\affiliation{
${}^{1}$Department of Mathematics and Physics, Graduate School of Science, Osaka City University, Osaka, Osaka 558-8585, Japan
\\
${}^{2}$Theory Center, Institute of Particle and Nuclear Studies, KEK, Tsukuba, Ibaraki 305-0801, Japan
}

%\date{\today}

\begin{abstract}
A supersymmetric solution to type II supergravity is constructed by superposing two hyper-K\"ahlers with torsion metrics.
The solution is given by a K\"ahler with torsion metric with $SU(3)$ holonomy. The metric is embedded into a heterotic solution obeying the Strominger system,
together with a Yang--Mills instanton obtained by the standard embedding. T dualities lead to an $SO(6)$ instanton describing a symmetry breaking from $E_8$ to $SO(10)$.
The compactification by taking a periodic array yields a supersymmetric domain wall solution of heterotic supergravity.
\end{abstract}


\preprint{OCU-PHYS 400}

\preprint{KEK-TH 1744}

%%%%%%%%%%%%%%%%%%%%%%%%%%%%%%%%%%%%%%%%%%%%%%%%%%%%%%%%%%%%%%%%%%%%%%%%%%%%%%%%%%%%%%

\maketitle

\section{Introduction}
The Green--Schwarz mechanism \cite{GSmechanism} 
is one of the cornerstones of superstring theory. 
Its role is twofold: First, of course, is to tell us 
how to cancel the gauge and gravitational anomalies of ten-dimensional 
type I and heterotic superstrings, which  
were apparently considered anomalous 
and hence unacceptable 
as consistent theories. 
With the mechanism, however, it turned out 
that all the anomalies were canceled out in a miraculous manner 
if and only if the gauge group was $SO(32)$ or $E_8\times E_8$,  
for the latter of which heterotic string theory has been constructed \cite{heterotic_string}.

The second important role of the Green--Schwarz mechanism is to constrain 
the background geometry through the modified Bianchi identity of 
the 3-form field $H$; 
the mechanism requires the 2-form $B$ field to vary under both 
the gauge and local Lorentz transformations so that the invariant 3-form 
field $H$ must be of the form
\begin{equation}\label{modifiledH}
H=d B-
\alpha' \left({\omega}_{3Y}
-{\omega}_{3L}^{-} \right),
\end{equation}
where 
${\omega}_{3Y}$ 
is 
the Chern--Simons 3-form associated with the Yang--Mills connection,
and ${\omega}_{3L}^{-}$
is also a Chern--Simons 3-form but made of a particular linear 
combination of the Levi-Civit$\acute{\text{a}}$ connection and the 3-form field:
\beqa\label{Hull}
\omega^-_{MAB}&=&\omega_{MAB}-H_{MAB}.
\eeqa
The equation (\ref{modifiledH}) leads to the Bianchi identity
\begin{equation}\label{Bianchi}
d H=\alpha' \left(tr \mathcal{F} \wedge \mathcal{F}-tr \mathcal{R}^{-} \wedge \mathcal{R}^{-} \right).
\end{equation}
This constrains the background geometry \cite{GSW} 
in such a way that the second 
Chern class of the gauge bundle be equal to the first Pontryagin class of 
the tangent bundle including torsion as in (\ref{Hull}).

Note that the combination (\ref{Hull}) 
is different from the one that appears 
in the supersymmetry(SUSY) variation of the gravitino
\beqa
\delta \psi_M&\propto  &\nabla^+ \varepsilon,
\label{gravitino_variation}
\eeqa
where $\nabla^+$ is the covariant derivative associated with 
the combination
\beqa
\omega^+_{MAB}&=&\omega_{MAB}+H_{MAB}.
\eeqa
The relevance of the difference between the two connections was 
pointed out by Bergshoeff and de Roo \cite{BdR}, and later 
emphasized by e.g., Refs. \cite{KimuraYi,MS}.

For $E_8\times E_8$ heterotic string theory on a six-dimensional 
space $M^6$ without $H$ fluxes,
the Killing spinor equation arising from the 
vanishing gravitino variation (\ref{gravitino_variation}) constrains 
$M^6$ to have $SU(3)$ holonomy, that is, to be Calabi--Yau. 
On the other hand, for the Bianchi identity 
(\ref{Bianchi}) to be satisfied, the easiest and most common way is to 
set the $\omega^+$ connection, 
which is nothing but the spin (Levi-Civit$\acute{\text{a}}$) connection for $H=0$, 
to be equal to a part of the gauge connection.
This is called the standard embedding \cite{Witten_New_Issues}. 
In this case,  
a part of the gauge field background is required to be $SU(3)$, 
and the gauge symmetry is partially broken to the centralizer 
$E_6(\times E_8)$. 
This reduction of the gauge symmetry is one of the 
hallmarks of Calabi--Yau compactifications of heterotic string theory.

If, on the other hand, there is a nonzero $H$ field, then the vanishing
gravitino variation (\ref{gravitino_variation}) asserts that the 
linear combination $\omega^+_{MAB}=\omega_{MAB}+H_{MAB}$ 
belongs to $SU(3)$ but says nothing about the other linear combination 
$\omega^-_{MAB}=\omega_{MAB}-H_{MAB}$ \cite{Strominger,BdR,KimuraYi}. 
Thus $\omega^-_{MAB}$ 
is generically in $SO(6)$ on the six-dimensional space $M^6$, and 
the gauge symmetry is broken to a smaller subgroup $SO(10)$, 
which is more favorable from the point of view of applications to 
string phenomenology. Note that, in the presence of  $H$ fluxes, 
$SO(10)$ is achieved by the ``standard embedding", that is, by 
simply equating the modified spin connection $\omega^-_{MAB}$ 
with a part of the gauge connection. This is in striking contrast 
to the $H=0$ Calabi--Yau case, in which one needs the nonstandard 
embedding that requires complicated mathematical 
machinery \cite{Witten_New_Issues,GSW} involving the construction of 
stable holomorphic vector bundles.

However, for the smeared intersecting NS5-brane solution, which is 
obtained as a superposition of two smeared  symmetric 5-brane 
solutions \cite{CHS} and is one of the simplest SUSY heterotic supergravity 
solutions with $H$ fluxes in the six-dimensional space, not only 
$\omega^+$ but also $\omega^-$ happens to be in $SU(3)$, and 
therefore the unbroken gauge symmetry is still $E_6$. 
The reason for this can be traced back to the parity invariance of the symmetric 
5-brane solution; indeed, the sign of $H$ is a matter of convention, and 
the configuration after the sign flip $H\rightarrow -H$ still remains 
a solution of the heterotic supergravity.

In this paper, we construct a supersymmetric heterotic supergravity 
solution such that $\omega^+$ is in $SU(3)$ (and hence a SUSY solution)
but $\omega^-$ is {\em not}
, by superposing two 
hyper-K\"{a}hlers with torsion (HKT) geometries. 
As already pointed out 
in Ref. \cite{CHS}, one can obtain HKT geometries by conformally transforming 
hyper-K\"{a}hler geometries.
We choose
the Gibbons--Hawking space as
the starting point and apply a conformal transformation to obtain 
a HKT geometry. Since the Gibbons--Hawking space is not parity 
invariant, the $\omega^-$ connection of the resulting HKT space is 
in $SO(4)$ but not in $SU(2)$, though $\omega^+$ still belongs to $SU(2)$.

We then smear the harmonic functions to those of 
two dimensions and take a superposition of two such geometries.   
Because of our superposition ansatz, we are forced to 
set some of the entries of the metric to zero in order to 
satisfy the equations of motion. Consequently, we find that 
the $\omega^-$ holonomy of the superposed solution remains 
to be $SO(4)$. We also show that by T duality this solution turns 
into one with $SO(5)$ or $SO(6)$ $\omega^-$ holonomy.

We also take a two-dimensional periodic array of the
``intersecting HKT" solutions to get a compact six-dimensional 
solution. We find that the fundamental parallelogram of the 
two-dimensional periodic array is separated into distinct smooth 
regions bordered by codimension-1 singularity hypersurfaces,
hence the name ``supersymmetric domain wall."  This novel 
solution has some interesting  properties, as we will see below.

This paper is organized as follows. In Sec. II, we give a brief 
review of HKT geometries obtained by conformal transformations 
acting on four-dimensional hyper-K\"ahler spaces.
In Sec. III, we consider a superposition of HKT spaces to construct 
a six-dimensional K\"ahler with torsion (KT) space with special properties 
which serves as a supersymmetric 
solution of type II supergravity. 
In Sec. IV, we embed this geometry into heterotic supergravity theory 
and take T dualities. In Sec. V, we compactify this six-dimensional 
space by taking a periodic array and study some of its properties.
The final section presents the summary and conclusion.

\section{HKT geometry as a conformal transform}

We start with a four-dimensional HKT metric ${g}_{HKT}$
obtained as a conformal transform of a hyper-K\"ahler metric, where for the 
latter we specifically consider the Gibbons--Hawking (GH) metric ${g}_{GH}$,
 \beqa
 {g}_{HKT}=\Phi \, {g}_{GH}.
 \eeqa
 The GH metric is given by \cite{GH}
\begin{equation}\label{GH}
{g}_{GH} =\frac{1}{\phi}\left(d\tau-\sum_{i=1}^3 \psi_i dx^i \right)^2
+\phi \sum_{i=1}^3 (dx^i)^2,
\end{equation}
where  $\phi$ and  $\textrm{\boldmath $\psi$}=(\psi_1, \psi_2, \psi_3)$ 
are scalar functions of the coordinates $ (x^1, x^2, x^3)$ of 
${\bf{R}}^3$ obeying the relation 
\beqa\label{phi_psi_relation}
\mbox{grad}\, \phi=\mbox{rot}\, \textrm{\boldmath $\psi$}.
\eeqa
$\Phi$ is a scalar field of which the properties will be described shortly. 
We define the orthonormal basis
\begin{equation}\label{base0}
{E}^{0}=\sqrt{\frac{\Phi}{\phi}}\left(d\tau-\sum_{i=1}^3 \psi_i dx^i \right),~{E}^{i}=\sqrt{\Phi\, \phi}\,dx^i~(i=1,2,3)
\end{equation}
so that the hypercomplex structure is given by the three complex structures 
${J}^a~(a=1,2,3)$ satisfying the quaternionic identities,
\beqa
{J}^a({E}^{\mu})=\bar{\eta}^a_{\mu \nu}{E}^{\nu},
\eeqa
where $\bar{\eta}^a_{\mu \nu}$ are the  't Hooft matrices.
The corresponding fundamental 2-forms are 
\beqa
{\Omega}^a=-\bar{\eta}^a_{\mu \nu}{E}^{\mu}\wedge {E}^{\nu}.
\eeqa

The HKT structure is defined by the 3-form torsion $T$ satisfying \cite{HP96}\cite{GP00}
\begin{equation}
T={J}^1 d {\Omega}^1={J}^2 d {\Omega}^2={J}^3 d {\Omega}^3.
\end{equation}
In the present case, 
we have
\begin{eqnarray}
& & T = -{E}_0 \log \Phi {E}^{123} + {E}_1 \log \Phi {E}^{023} + {E}_2 \log \Phi {E}^{031} \nonumber \\
& & ~~~~~~+{E}_3 \log \Phi {E}^{012}
\end{eqnarray}
in terms of dual vector fields ${E}_\mu$ to the 1-forms (\ref{base0}),
\begin{equation}
{E}_0=\sqrt{\frac{\phi}{\Phi}} \frac{\partial}{\partial \tau},~~
{E}_i=\frac{1}{\sqrt{\Phi\, \phi}}\left(\frac{\partial}{\partial x^i}+\psi_i \frac{\partial}{\partial \tau}\right)
\end{equation}
and
\beqa
{E}^{\mu \nu \lambda}={E}^{\mu}\wedge {E}^{\nu} \wedge {E}^{\lambda}.
\eeqa  
The exterior derivative is calculated as
\begin{equation}
d T=-\frac{1}{\Phi^2 \phi}\left(\sum_{\mu=0}^3 {V}_\mu^2 \Phi \right){E}^{0123}
\end{equation}
with the vector fields ${V}_\mu=\sqrt{\Phi \,\phi}\,{E}_\mu$. Therefore, if $\Phi$ is chosen to be a harmonic function 
with respect to the GH metric (\ref{GH}), 
then 
the torsion $T$ becomes a closed 3-form.

Using this 
$T$, we introduce the 
 two types of connections $\nabla^{\pm}$,
\begin{equation}\label{BH}
\nabla ^{\pm}_{X}Y = \nabla_{X}Y \pm \frac{1}{2}\sum_{\mu=0}^{3}T(X,Y,E_{\mu})E_{\mu}, \,
\end{equation}
where 
$\nabla$ is a Levi-Civit\'a connection. 
The corresponding connection 1-forms $\mathcal{\omega}^{\pm \,\mu}_{~~~\nu}$ are defined by
\begin{equation}
\nabla ^{\pm}_{E_\mu}{E}_\nu=\mathcal{\omega}^{\pm \,\lambda}_{~~~\nu}(E_\mu){E}_\lambda,
\end{equation}
and the curvature 2-forms are written as 
\begin{equation}
\mathcal{R}^{\pm \, \mu}_{~~~\nu}=d \mathcal{\omega}^{\pm \,\mu}_{~~~\nu} + \mathcal{\omega}^{\pm \, \mu} _{~~~\lambda}
 \wedge \mathcal{\omega}^{\pm \, \lambda}_{~~~\nu}.
\end{equation}
The torsion curvature $\mathcal{R}^{+ \, \mu}_{~~~\nu}$ satisfies 
the $SU(2)$ holonomy condition
\begin{equation}
\mathcal{R}^{+}_{~~01}+\mathcal{R}^{+}_{~~23}=0,~\mathcal{R}^{+}_{~~02}+\mathcal{R}^{+}_{~~31}=0,~
\mathcal{R}^{+}_{~~03}+\mathcal{R}^{+}_{~~12}=0.
\end{equation}
On the other hand, if the torsion $T$ is a closed 3-form, that is, 
$\Phi$ is a harmonic function, then the curvature $\mathcal{R}^{- \, \mu}_{~~~\nu}$ becomes an anti self dual 2-form,
which may be regarded as a Yang--Mills instanton with the gauge group  $SU(2)\times SU(2)=SO(4)$.

 \section{Intersecting HKT metrics}
 
 In the previous section we have seen that the HKT metrics obtained by a 
 conformal transformation have 
  $\mathcal{\omega}^{+ \,\mu}_{~~~\nu}$ 
  in $SU(2)$ but $\mathcal{\omega}^{- \,\mu}_{~~~\nu}$ 
 in $SO(4)$ strictly larger than $SU(2)$ as long as the original GH space 
 is not a flat Euclidean space.
In this section we construct their six-dimensional analogs by 
superposing two such HKT metrics embedded in different 
four-dimensional subspaces. 
This construction is motivated by that used in constructing 
intersecting brane solutions \cite{AEH,O}\footnote{The term ``intersecting'' 
in the (commonly used) name 
is misleading since they are smeared and hence do not have intersections 
with larger codimensions. See, e.g., Ref. \cite{MR} for recent developments 
in constructing localized intersecting brane solutions in supergravity.};
namely, we assume the form of the metric as
\begin{eqnarray}\label{intersect}
& & g = \Phi \tilde{\phi}\phi \tilde{\phi}((dx^1)^2 + (dx^2)^2) + \Phi \phi (dx^3)^2 \nonumber \\
& & ~~~~+ \frac{\Phi}{\phi}(dx^4-\psi dx^3)^2 + \tilde{\Phi}\tilde{\phi}(dx^5)^2 + \frac{\tilde{\Phi}}{\tilde{\phi}}(dx^6 - \tilde{\psi}dx^5)^2.\label{INS5Bmetric}\nonumber \\
\end{eqnarray}

The HKT metric that we have considered in the previous section 
is characterized by a triplet $(\Phi, \phi, \psi)$ on ${\bf{R}}^3=\{ (x^1, x^2, x^3) \}$ obeying 
(\ref{phi_psi_relation}). 
So at first it might seem that $(\Phi, \phi)$ or $(\tilde{\Phi}, \tilde \phi)$ could to be functions of 
$(x^1, x^2, x^3)$ or $(x^1, x^2, x^5)$, 
and $dx^4-\psi dx^3$ or $dx^6-\tilde{\psi}dx^5$ could be replaced with 
a more general form $dx^4-\sum_{i=1,2,3}\psi_i dx^i$ or $dx^6-\sum_{i=1,2,5}\tilde\psi_i dx^i$, respectively.
However, it turns out that such a more general ansatz does not lead 
to a metric with $SU(3)$ holonomy even in the case $\Phi=\tilde \Phi =1$.
Thus we are led to consider the metric of the form (\ref{INS5Bmetric}), assuming the following: 
\begin{itemize}
\item $(\Phi, \phi)$ and $(\tilde{\Phi}, \tilde{\phi})$ are harmonic functions 
on the two-dimensional flat space ${\bf{R}^2}=\{(x^1, x^2)\}$.
\item 
$\textrm{\boldmath $\psi$}=(0, 0, \psi)$ and $\textrm{\boldmath $\tilde{\psi}$}=(0, 0, \tilde{\psi})$
, of which the components are harmonic functions on ${\bf{R}^2}$ satisfying the Cauchy--Riemann conditions
\begin{eqnarray}
\frac{\partial \phi}{\partial x_2}&=&-\frac{\partial \psi}{\partial x_1},~~~\frac{\partial \phi}{\partial x_1}=\frac{\partial \psi}{\partial x_2}, \nonumber\\
\frac{\partial \tilde{\phi}}{\partial x_2}&=&-\frac{\partial \tilde{\psi}}{\partial x_1},~~~\frac{\partial \tilde{\phi}}{\partial x_1}=\frac{\partial \tilde{\psi}}{\partial x_2}.
\end{eqnarray}
\end{itemize}
Under these assumptions, we will show that
a six-dimensional space $M^6$ with the metric (\ref{intersect}) has the following KT structure:
\begin{itemize}
\item[(a)] a closed Bismut torsion [see Eq. (\ref{torsion})], 
\item[(b)] an exact Lee form [see Eq. (\ref{dilaton})],
\item[(c)] a Bismut connection $\nabla ^{+}$ with $SU(3)$ holonomy [see Eq. (\ref{holonomy})].
\end{itemize}

We first introduce an orthonormal basis
\begin{eqnarray}
& & {e}^{1} = \sqrt{\Phi\tilde{\Phi}\phi\tilde{\phi}}dx^{1} \;,\; {e}^{2} = \sqrt{\Phi\tilde{\Phi}\phi\tilde{\phi}}dx^{2} \,, \nonumber \\
& & {e}^{3} = \sqrt{\Phi\phi}dx^{3} \;,\; {e}^{4} = \sqrt{\frac{\Phi}{\phi}}(dx^{4} - \psi dx^{3}) \,, \nonumber \\
& & {e}^{5} = \sqrt{\tilde{\Phi}\tilde{\phi}}dx^{5} \;,\; {e}^{6} = \sqrt{\frac{\tilde{\Phi}}{\tilde{\phi}}}(dx^{6} - \tilde{\psi}dx^{5}) \,. \label{base1}
\end{eqnarray}
The space $M^{6}$ has a natural complex structure $J$ defined by
\begin{equation}
J({e}^1) = \epsilon_{1}{e}^{2} \,,\, J({e}^3) = \epsilon_{2}{e}^{4} \,,\, J({e}^5) = \epsilon_{3}{e}^{6}\,.
\end{equation}
Indeed, it is easy to see that the Nijenhuis tensor associated with $J$ vanishes under the condition $|\epsilon_i|=1 (i=1,2,3)$ and $\epsilon_1\epsilon_2=\epsilon_1\epsilon_3=-1$.
Then, the metric (\ref{intersect}) becomes Hermitian with respect to the complex structure $J$, and
the fundamental 2-form $\kappa$ takes the form
\begin{equation}\label{kahler}
\kappa = \epsilon_{1}{e}^{1} \wedge {e}^{2} + \epsilon_{2}{e}^{3} \wedge {e}^{4} + \epsilon_{3}{e}^{5} \wedge {e}^{6} \,.
\end{equation}
The Bismut torsion $T$ is uniquely determined by
\begin{equation}
\nabla ^{+}_X g = 0 \;,\; \nabla ^{+}_X \kappa = 0 \,.
\end{equation}
Explicitly we have 
\begin{eqnarray}\label{torsion}
& & T = -J\,d \kappa \nonumber\\
& &   ~~= \frac{1}{\Phi\sqrt{\Phi\tilde{\Phi}\phi\tilde{\phi}}}(\partial _{1}\Phi e^{234}-\partial _{2}\Phi e^{134}) \nonumber \\
& &       ~~~~~~~+ \frac{1}{\tilde{\Phi}\sqrt{\Phi\tilde{\Phi}\phi\tilde{\phi}}}(\partial _{1}\tilde{\Phi}e^{256}-\partial _{2}\tilde{\Phi}e^{156}).
\end{eqnarray}
It should be noticed that in our case the Bismut torsion is a closed 3-form, $dT = 0$. 
We shall refer to $\nabla ^{+}$ and $\nabla ^{-}$ as the Bismut connection and Hull connection, respectively, according to Ref. \cite{MS}.
The Lee form $\theta$ is a 1-form defined by
$ \theta = -J \delta \kappa$ \cite{IP}, which becomes a closed 1-form,
\begin{equation}\label{dilaton}
\theta=2d \varphi,~~~\varphi=\log \sqrt{\Phi \tilde{\Phi}}.
\end{equation}

We will identify the Bismut torsion with 3-form flux, $T= H$, and the function $\varphi$ with a dilaton.
It is shown that the Ricci form \cite{IP} of the Bismut connection vanishes, which is equivalent to the condition
\begin{equation}\label{holonomy}
\epsilon_1 \mathcal{R}^{+}_{~12}+\epsilon_2 \mathcal{R}^{+}_{~34}+\epsilon_3 \mathcal{R}^{+}_{~56}=0,
\end{equation}
so that the holonomy of $\nabla^+$ is contained in $SU(3)$ and $M^{6}$ admits two independent Weyl Killing spinors obeying $\nabla^{+}_X \varepsilon=0$ in type II theory.
Thus the triplet $(g, H, \varphi)$ gives rise to a supersymmetric solution to the type II supergravity theory.

\section{Embedding into heterotic string theory and T duality}
We study supersymmetric solutions describing heterotic flux compactification.  
The bosonic part of the string frame action, up to the first order in the $\alpha$' expansion,  is given by
\begin{eqnarray}\label{heterotic action}
S & = & \frac{1}{2 \kappa^2}\int d^{10}x \sqrt{-g}e^{-2 \varphi}\left( R + 4(\nabla \varphi)^{2} \right.\nonumber \\
  &   & - \frac{1}{12}H_{MNP}H^{MNP} \nonumber \\
  &   &     \left. - \alpha'(tr\mathcal{F}_{MN}\mathcal{F}^{MN} - tr\mathcal{R}^{-}_{MN}\mathcal{R}^{-MN}) \right) .
\end{eqnarray}
It is assumed that ten-dimensional spacetimes take the form $R^{1,3} \times M^6$, where $M^6$ is a
six-dimensional space admitting a Killing spinor $\varepsilon$,
\begin{equation}\label{instanton}
\nabla^{+}_a \varepsilon=0,~~
\left(\gamma^a \partial_a \varphi+\frac{1}{12}H_{abc}\gamma^{abc} \right)\varepsilon=0,~~
\mathcal{F}_{ab}
\gamma^{ab}\varepsilon=0.
\end{equation}
This system together with the anomaly cancellation condition
\begin{equation}\label{anomaly}
d H=\alpha' \left(tr \mathcal{F} \wedge \mathcal{F}-tr \mathcal{R}^{-} \wedge \mathcal{R}^{-} \right)
\end{equation}
is known as the Strominger system \cite{Strominger}.

Now, we turn to the heterotic solution obeying the Strominger system.
If the curvature $\mathcal{R}^-$ in the anomaly condition (\ref{anomaly}) is given by the Hull connection $\nabla^-$, 
we can choose a non-Abelian gauge field as $\mathcal{F}=\mathcal{R}^-$ since the 3-form flux (\ref{torsion}) is closed by the identification $T=H$.
This is a form of the usual standard embedding.
Combining the well-known identity 
\begin{equation}
\mathcal{R}^{+}_{abcd}-\mathcal{R}^{-}_{cdab}=\frac{1}{2}(dT)_{abcd}=0
\end{equation}
with the holonomy condition (\ref{holonomy}),
we can see that the gauge field $\mathcal{F}$ is an instanton satisfying the third equation in (\ref{instanton}).

Apparently, $\mathcal{F}$ seems to take values in $SO(6)$ $\subset$ $E_8$,
which would describe a symmetry breaking from $E_8$ to $SO(10)$.
However, 
for generic choices of 
the harmonic functions $\phi$, $\Phi$, $\tilde{\phi}$, and $\tilde{\Phi}$, 
it is not ensured that
the metric (\ref{INS5Bmetric}) can remain non-negative, and 
the dilaton (\ref{dilaton}) can remain real valued.
Therefore, to get a meaningful solution we are forced to impose
\beqa
\phi=\tilde{\phi}=\Phi=\tilde{\Phi}.
\eeqa
With this condition, the holonomy of $\nabla^+$ remains $SU(3)$, but the instanton $\mathcal{F}$ reduces 
to a proper Lie subalgebra $SO(4)$ of $SO(6)$, and 
the centralizer is $SO(12)$.

To recover the $SO(6)$ instanton, 
we apply a T-duality transformation. 
From (\ref{intersect}), (\ref{torsion}), and (\ref{dilaton}), 
with $\phi=\tilde{\phi}=\Phi=\tilde{\Phi}$, 
we have the following metric with
$SU(3)$ holonomy , 3-form flux, and dilaton:
\begin{eqnarray}
g & = & \phi^{4}((dx^{1})^{2} + (dx^{2})^{2}) + \phi^{2}(\, (dx^{3})^{2} + (dx^{5})^{2} \,) \nonumber\\
  & + & (dx^{4} - \psi dx^{3})^{2} + (dx^{6} - \psi dx^{5})^{2}, \label{metric_all_phi's_equal} \\ 
H & = & - \frac{1}{\phi^{3}}( \partial _{2}\phi e^{134} - \partial _{1}\phi e^{234} + \partial _{2}\phi e^{156} - \partial _{1}\phi e^{256} ) , \nonumber \\
\label{H_all_phi's_equal}\\
\varphi &=& \log{|\phi|}.\label{dilaton_all_phi's_equal}
\end{eqnarray}
The metric (\ref{metric_all_phi's_equal}) has isometries $U(1)^4$ generated by Killing vector fields $\partial_a~~(a=3,4,5,6)$.
Therefore, we can T dualize the type II solution $(g, H, \varphi)$ along directions of these isometries.
It is easy to see that the solution is inert under the T duality along $x^4$ and $x^6$; 
the T dualities along the remaining directions give nontrivial 
deformations of the solutions, preserving one-quarter of supersymmetries.\footnote{See, e.g., Ref. \cite{P} for the classification of supersymmetric solutions to heterotic supergravity.}

We first T dualize the solution along $x^3$. 
The resulting solution  $(\hat{g},\hat{H},\hat{\varphi})$ is given by
\begin{eqnarray}\label{T1metric}
\hat{g} & = & \phi^{4}(\, (dx^{1})^{2} + (dx^{2})^{2} \,)+\frac{1}{\phi^{2} + \psi^{2}}( d\hat{x}^{3} + \psi dx^{4} )^{2} \nonumber \\
        &   &  + \frac{\phi^{2}}{\phi^{2} + \psi^{2}}(dx^{4})^{2} + \phi^{2}(dx^{5})^{2} + (dx^{6} - \psi dx^{5})^{2}\,,\\
\label{T1flux}
\hat{H} & = & \frac{1}{\phi^{3}(\phi^{2} + \psi^{2})}((\phi^{2} + \psi^{2})\partial _{2}\phi + 2\psi(\phi\partial _{1}\phi - \psi\partial _{2}\phi) \,)\hat{e}^{134} \nonumber \\
        &   &  - \frac{1}{\phi^{3}(\phi^{2} + \psi^{2})}(\, (\phi^{2} + \psi^{2})\partial _{1}\phi - 2\psi(\phi\partial _{2}\phi + \psi\partial _{1}\phi) \,)\hat{e}^{234} \nonumber\\
        &   &    - \frac{1}{\phi^{3}}( \partial _{2}\phi \hat{e}^{156} - \partial _{1}\phi \hat{e}^{256})\,,\\
\label{T1dilaton}
\hat{\varphi} & = & \frac{1}{2}\log\left({\frac{1}{(\phi^{2} + \psi^{2})}\phi^{2}}\right) \,.
\end{eqnarray}
Here, the orthonormal basis is defined by
\begin{eqnarray}
& & \hat{e}^{1} = \phi^{2}dx^{1} \;,\; \hat{e}^{2} = \phi^{2}dx^{2}, \nonumber \\
& & \hat{e}^{3} = \frac{1}{\sqrt{\phi^{2} + \psi^{2}}}\left ( d\hat{x}^{3} + \psi dx^{4} \right ) \;,\; \hat{e}^{4} = \frac{\phi}{\sqrt{\phi^{2} + \psi^{2}}}dx^{4}, \nonumber \\
& & \hat{e}^{5} = \phi dx^{5} \;,\; \hat{e}^{6} = dx^{6} - \psi dx^{5} \,. \label{new orthonormal base1}
\end{eqnarray}
Then, we have a deformed complex structure $\hat{J}$,
\begin{equation}
\hat{J}\hat{e}_{1} = \epsilon_{1}\hat{e}_{2} \;,\; 
\hat{J}\hat{e}_{3} = \epsilon_{2}\hat{e}_{4} \;,\; 
\hat{J}\hat{e}_{5} = \epsilon_{3}\hat{e}_{6} \,
\end{equation}
with $|\epsilon_i|=1 (i=1,2,3)$ and $\epsilon_1\epsilon_2=\epsilon_1\epsilon_3=-1$.
The associated fundamental two-form $\hat{\kappa}$ takes the same form as (\ref{kahler}), and the Bismut connection $\nabla^+$ has an $SU(3)$ holonomy.
In this case, 
it turns out that 
the Hull connection $\nabla^-$ is in $SO(5)$, which is still smaller than $SO(6)$.

Thus, we further T dualize the solution  $(\hat{g},\hat{H},\hat{\varphi})$ once more along $x^5$ 
and finally obtain $(\tilde{g},\tilde{H},\tilde{\varphi})$:
\begin{eqnarray}\label{T2metric}
& & \tilde{g} = \phi^{4}(\, (dx^{1})^{2} + (dx^{2})^{2} \,)+ \frac{1}{\phi^{2} + \psi^{2}}(d\hat{x}^{3} + \psi dx^{4})^{2} \nonumber \\
& & ~           +\frac{1}{\phi^{2} + \psi^{2}}( d\tilde{x}^{5} + \psi dx^{6} )^{2} + \frac{\phi^{2}}{\phi^{2} + \psi^{2}}((dx^{4})^{2} +(dx^{6})^{2}), \nonumber \\
\end{eqnarray}  
\begin{eqnarray}\label{T2flux}                             
\tilde{H} = \frac{1}{\phi^{3}(\phi^{2} + \psi^{2})}( (\phi^{2} + \psi^{2})\partial _{2}\phi + 2\psi(\phi\partial _{1}\phi - \psi\partial _{2}\phi) )\tilde{e}^{134} \nonumber \\
            - \frac{1}{\phi^{3}(\phi^{2} + \psi^{2})}(\, (\phi^{2} + \psi^{2})\partial _{1}\phi - 2\psi(\phi\partial _{2}\phi + \psi\partial _{1}\phi) )\tilde{e}^{234} \nonumber \\
              + \frac{1}{\phi^{3}(\phi^{2} + \psi^{2})}( (\phi^{2} + \psi^{2})\partial _{2}\phi + 2\psi(\phi\partial _{1}\phi - \psi\partial _{2}\phi) )\tilde{e}^{156} \nonumber \\
               - \frac{1}{\phi^{3}(\phi^{2} + \psi^{2})}( (\phi^{2} + \psi^{2})\partial _{1}\phi - 2\psi(\phi\partial _{2}\phi + \psi\partial _{1}\phi) )\tilde{e}^{256}, \nonumber \\
\end{eqnarray}
\begin{equation}\label{T2dilaton}
\tilde{\varphi} = \frac{1}{2}\log\left(\frac{1}{(\phi^{2} + \psi^{2})(\phi^{2} + \psi^{2})}\phi^{2}\right) \,.
\end{equation}
The orthonormal basis is defined by
\begin{eqnarray}
& & \tilde{e}^{1} =\hat{e}^{1}  \;,\; \tilde{e}^{2} = \hat{e}^{2} \,,
\, \tilde{e}^{3} = \hat{e}^{3}  \;,\;  \tilde{e}^{4} = \hat{e}^{4} \, \nonumber \\
& & \tilde{e}^{5} = \frac{1}{\sqrt{\phi^{2} + \psi^{2}}}(d\tilde{x}^{5} + \psi dx^{6} ) \;,\; \tilde{e}^{6}  = \frac{\phi}{\sqrt{\phi^{2} + \psi^{2}}}dx^{6} \,. \nonumber \\
\end{eqnarray}
In this basis the complex structure $\tilde{J}$ is given by
\begin{equation}
\tilde{J} \tilde{e}^{1}= \epsilon_{1}\tilde{e}^{2} \;,
\; \tilde{J}\tilde{e}^{3} = \epsilon_{2}\tilde{e}^{4} \;,
\; \tilde{J}\tilde{e}^{4} = \epsilon_{3}\tilde{e}^{2} \,
\end{equation}
with $|\epsilon_i|=1 (i=1,2,3)$ and $\epsilon_1\epsilon_2=\epsilon_1\epsilon_3=-1$. 
It can be verified that this solution has an $SU(3)$ Bismut connection $\nabla^+$ and
$SO(6)$ Hull connection $\nabla^-$ as desired.\\

\section{SUSY domain wall metric}

The last topic concerns the construction of type II/heterotic supersymmetric solutions 
on a {\em compact} six-dimensional 
space with the Hull connection not being in $SU(3)$. 
Since the triples obtained in the previous section
depend only on $x^1$ and $x^2$, we can 
compactify the $x^3$, $x^4$, $x^5$ and $x^6$ 
spaces on $T^4$ by simply identifying periodically, whereas we consider a periodic array of 
copies of the solution along the $x^1$ and $x^2$ directions.

Let us consider a periodic array of $(g,H,\varphi)$
[Eqs. (\ref{metric_all_phi's_equal}), (\ref{H_all_phi's_equal}), and (\ref{dilaton_all_phi's_equal})], 
$(\hat{g},\hat{H},\hat{\varphi})$
[(\ref{T1metric}), (\ref{T1flux}), and (\ref{T1dilaton})], or
$(\tilde{g},\tilde{H},\tilde{\varphi})$
[(\ref{T2metric}), (\ref{T2flux}), and (\ref{T2dilaton})], 
which are characterized by a pair of harmonic functions $\phi$ and $\psi$.
In two dimensions both the real and imaginary parts of any holomorphic function are harmonic. 
Thus we can take $\phi$ to be, say, the real part of 
any doubly periodic, holomorphic function. In this case, $\psi$ may be taken to be 
the imaginary part of the same doubly periodic function.

\begin{figure}[h]%
\centering
\includegraphics[height=0.25\textheight]{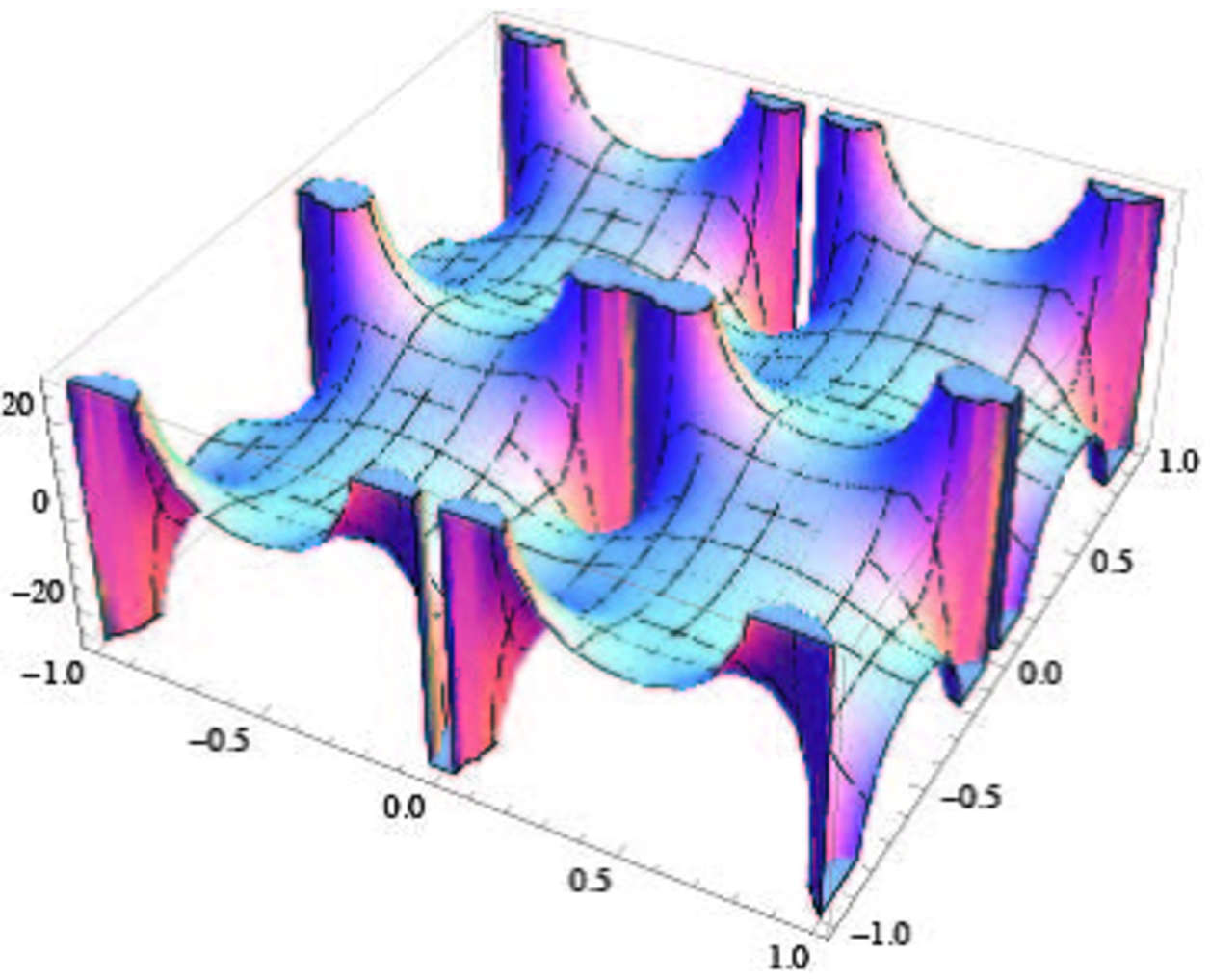}
~
\includegraphics[height=0.25\textheight]{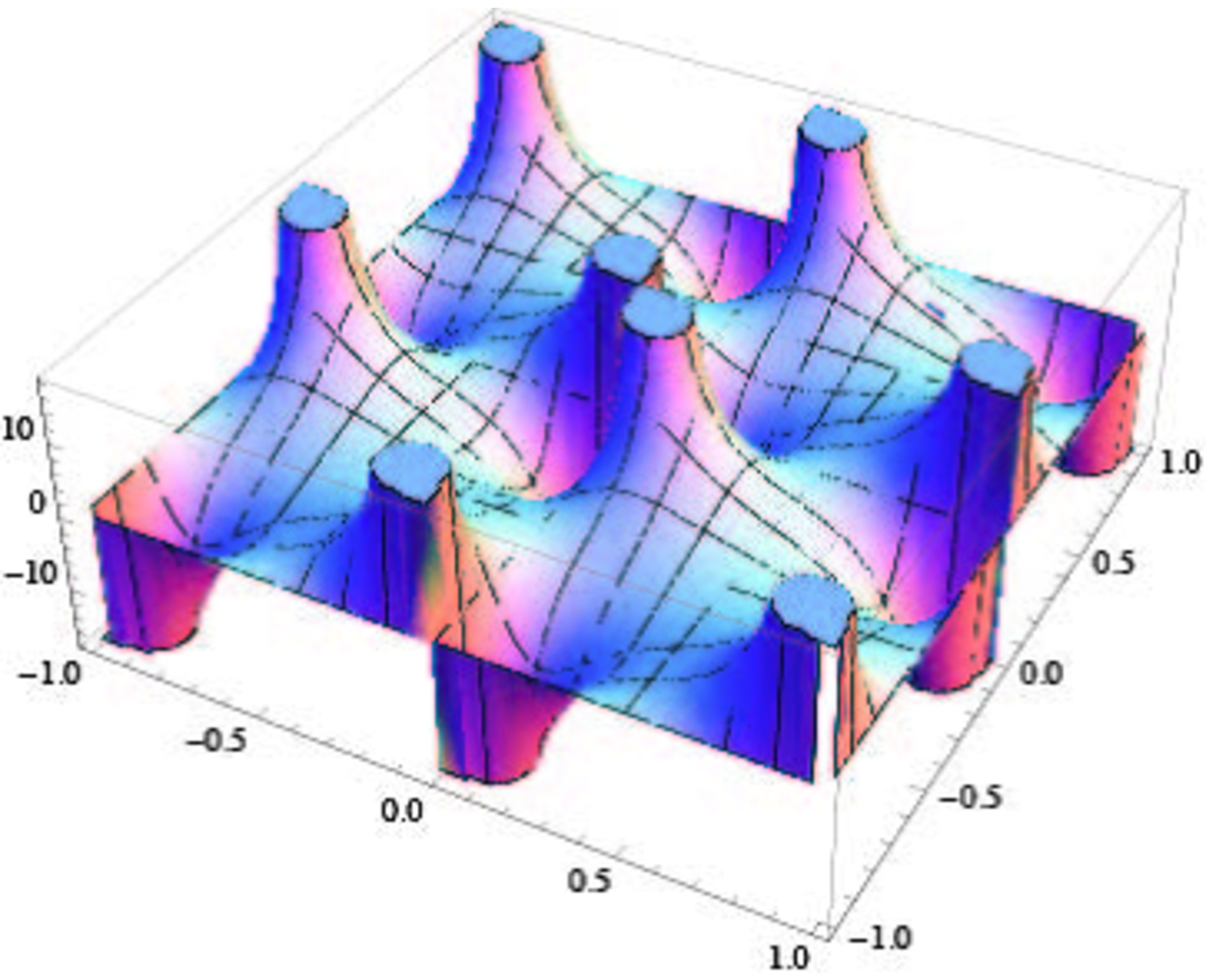}
\caption{\label{pe}  The real (upper plot) and imaginary (lower plot) parts of the $\wp$ 
function. The fundamental parallelogram can be taken to be 
$-\frac12\leq \frac{x^1}l \leq \frac12$ and $-\frac12\leq \frac{x^2}l \leq \frac12$.}
\end{figure}

Since the only nonsingular holomorphic function on $T^2$ is a constant function, 
we need to allow some pole singularities in the fundamental parallelogram of 
the periodic array,
which may be seen to be in accordance with the no-go theorems against smooth 
flux compactifications \cite{GKP,KimuraYi}. The doubly periodic meromorphic 
functions are known as elliptic functions. It is well known that, for a given periodicity, 
the field of elliptic functions is generated by Weierstrass's $\wp$ function and its derivative  
$\wp'$. In the following, we consider, as a typical example, the compactification 
of 
$(g,H,\varphi)$,
$(\hat{g},\hat{H},\hat{\varphi})$,
and 
$(\tilde{g},\tilde{H},\tilde{\varphi})$
on a square torus of side $l$ by taking 
\beqa
\phi(x^1,x^2)
&=&\mbox{Re}~{\wp}(z),\\
\psi(x^1,x^2)
&=&\mbox{Im}~{\wp}(z),
\eeqa
where $\wp(z)$ is of modulus $\tau=i$ or 
$\tau=e^{\frac{\pi i}5}$
and $z=l^{-1}(x^1+ i x^2)$.
Our solutions are determined entirely by Weierstrass's $\wp $ function without any reference to $\alpha'$ because of the choice $\mathcal{F} = \mathcal{R}^{-}$ that causes the rhs of (\ref{anomaly}) to be closed.
Note that they solve the heterotic equations of motion up to $\mathcal{O}(\alpha')$.

\begin{figure}[h]%
\centering
\includegraphics[height=0.25\textheight]{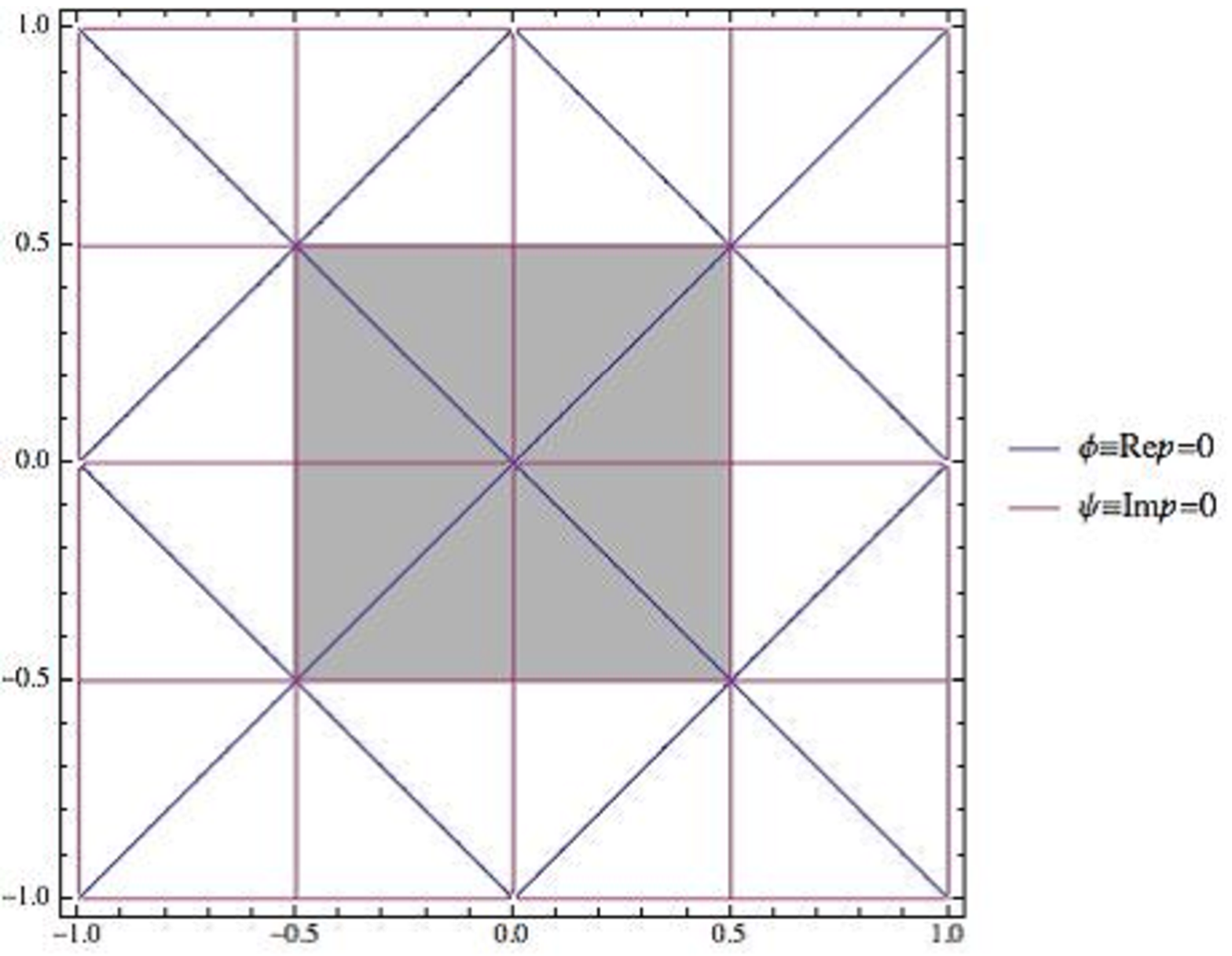}
~
\includegraphics[height=0.25\textheight]{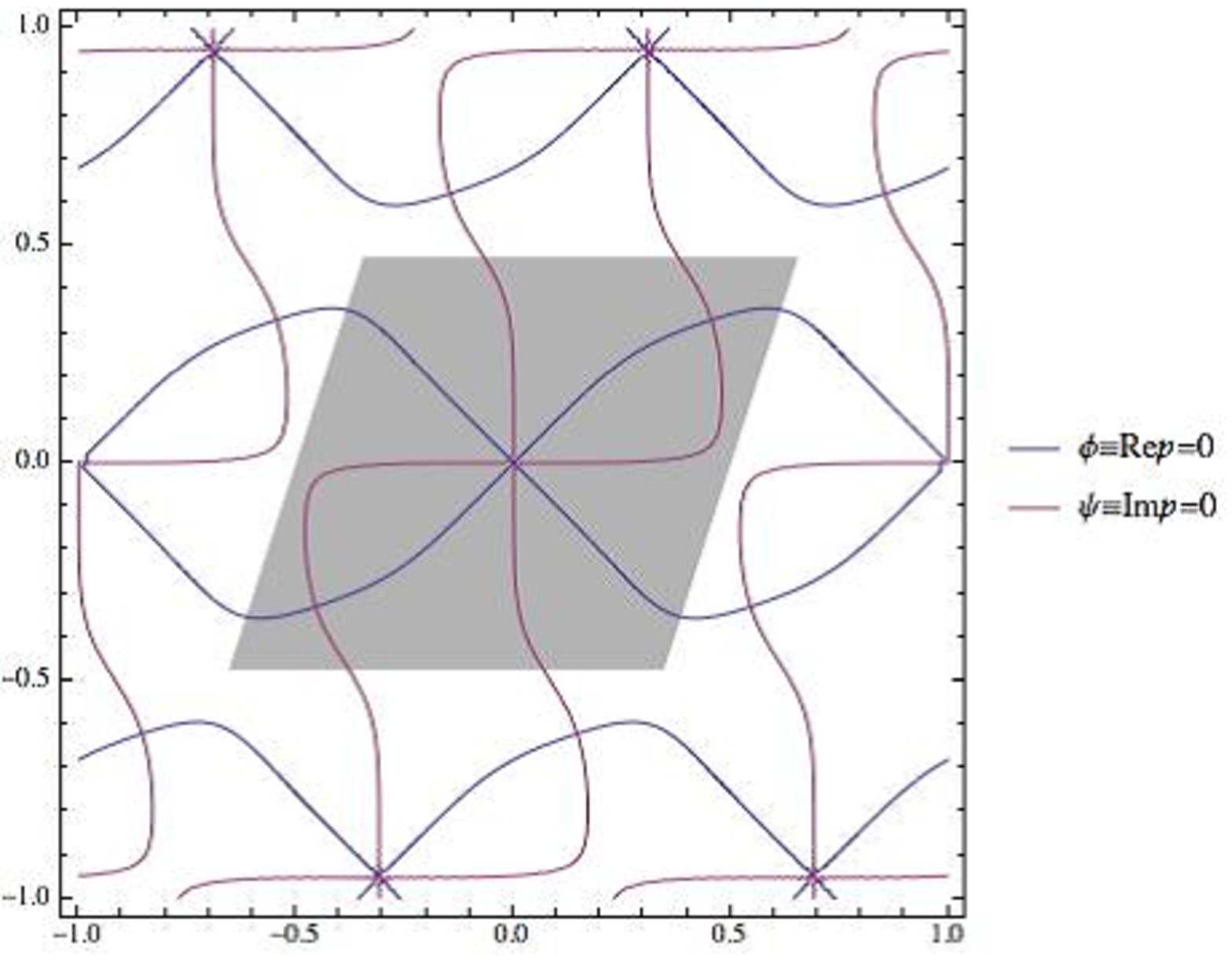}
\caption{\label{pe}  The zero loci of the real and imaginary parts of the $\wp$ 
function for the modulus $\tau=i$ (upper plot) and $\tau=e^{\frac{\pi i}5}$ (lower plot).
The shaded region is the fundamental parallelogram.  }
\end{figure}

The real and imaginary parts of $\wp(z)$ are shown in Fig. 1. 
We see that $\phi$ may take negative as well as positive values, but note that 
the metric (\ref{metric_all_phi's_equal}), (\ref{T1metric}), or (\ref{T1metric}) 
depends on $\phi$ through $\phi^2$ as we designed, 
so the solution is only singular 
where $\phi$ vanishes (as well as $\phi$ diverges).
Also, negative $\psi$ causes no problem as long as $\phi$ is nonzero.

For any case of $(g,H,\varphi)$,
$(\hat{g},\hat{H},\hat{\varphi})$,
or
$(\tilde{g},\tilde{H},\tilde{\varphi})$, 
some of the components of the metric vanish where $\phi=0$, and hence 
the solution is singular. Also, the ``string coupling'' (= exponential of the dilaton) 
vanishes there.  The $\phi=0$ curves are shown in Fig. 2 for the cases 
$\tau=i$ and $\tau=e^{\frac{\pi i}5}$. 
For both cases, 
we see that the fundamental parallelogram (shown by the shaded region)  
is separated into two distinct smooth 
regions bordered by the codimension-1 singularity hypersurfaces.
The two singularity hypersurfaces intersect at $x^1=x^2=0$, where 
the $\wp$ function has a unique double pole; its real and imaginary 
parts rapidly fluctuate at $x^1=x^2=0$. 
More details about the solution will be reported elsewhere.

\section{Conclusions}
In this paper we have shown that 
two HKT metrics given by $(\Phi, \phi, \textrm{\boldmath $\psi$})$ and $(\tilde{\Phi}, \tilde{\phi}, \tilde{\textrm{\boldmath $\psi$}})$ 
can be superposed and 
lifted to a six-dimensional smeared intersecting solution of type II supergravity 
if the functions $\Phi , \phi, \tilde{\Phi}$, and $ \tilde{\phi}$ are restricted to harmonic functions 
on the two-dimensional flat space ${\bf{R}^2}=\{(x^1, x^2)\}$,
together with 
$\textrm{\boldmath $\psi$}=(0, 0, \psi)$ and $\tilde{\textrm{\boldmath $\psi$}}=(0, 0, \tilde{\psi})$
satisfying the Cauchy--Riemann conditions. 
The simplest geometry that we have considered has an $SO(4)$ $\nabla^-$ connection 
that leads to the $SO(10)$ unbroken gauge symmetry if it is embedded to 
heterotic string theory as an internal space. By T-duality transformations 
we have obtained one having an $SO(5)$ or $SO(6)$ $\nabla^-$ holonomy.
We have also compactified this six-dimensional KT space by taking a periodic 
array to find a supersymmetric domain wall solution of heterotic supergravity 
in which the fundamental parallelogram of the 
two-dimensional periodic array is separated into distinct smooth 
regions bordered by codimension-1 singularity hypersurfaces. 
It would be interesting to solve the gaugino Dirac equation on this background 
and compare the spectrum with the corresponding $E_8$-type supersymmetric nonlinear 
sigma model \cite{IrieYasui}, similarly to what has been done in the $SU(3)$ 
$\nabla^-$ case \cite{MY}.

\section*{Acknowledgements}

Y.Y. is supported by the Grant-in-Aid for Scientific Research Grant No.~23540317,
and S.M. is supported by Grant No.~25400285 from
The Ministry of Education, Culture, Sports, Science
and Technology of Japan.

\end{document}